\documentclass[conference]{IEEEtran}
 \usepackage[pdftex]{graphicx}

\usepackage{amsmath,amsthm}
\usepackage{amsmath,bm}
\usepackage{amsfonts}
\usepackage{cases}
\usepackage{subfigure}
\usepackage{booktabs}

\ifCLASSINFOpdf
\else
\fi
\begin{document}
\title{Spectrum sharing in energy harvesting cognitive radio networks: A cross-layer perspective}

%\author{\IEEEauthorblockN{Michael Shell}
%\IEEEauthorblockA{School of Electrical and\\Computer Engineering\\
%Georgia Institute of Technology\\
%Atlanta, Georgia 30332--0250\\
%Email: http://www.michaelshell.org/contact.html}
%\and
%\IEEEauthorblockN{Homer Simpson}
%\IEEEauthorblockA{Twentieth Century Fox\\
%\and
%\IEEEauthorblockN{James Kirk\\ and Montgomery Scott}
%\IEEEauthorblockA{Starfleet Academy\\
%San Francisco, California 96678--2391\\

\author{\IEEEauthorblockN{Tian Zhang\IEEEauthorrefmark{1}\IEEEauthorrefmark{2} and
Wei Chen\IEEEauthorrefmark{1}
}
\IEEEauthorblockA{\IEEEauthorrefmark{1}
Department of Electronic Engineering, Tsinghua University, Beijing 100084, China
}
\IEEEauthorblockA{
\IEEEauthorblockA{\IEEEauthorrefmark{2}School of Information Science and Engineering,
Shandong Normal University, Jinan 250100, China }
\\ Email: tianzhang.ee@gmail.com, wchen@tsinghua.edu.cn}
}


%Telephone: (800) 555--1212, Fax: (888) 555--1212}


\maketitle

\begin{abstract}
In the paper, we present a cross-layer perspective on data transmission in energy harvesting cognitive radio networks (CRNs). The delay optimal power allocation is studied while taking into account the randomness of harvested energy, data generation, channel state and the grid price. To guarantee primary user (PU)'s transmission, its Signal-Interference-Ratio (SIR) should be no less than a threshold. Each user, including PU as well as secondary user (SU), has energy harvesting devices, and the PU can also purchases the grid power. Each user is rational and selfish to minimize its own the buffer delay. We formulate a stochastic Stackelberg game in a bilevel manner. After decoupling via rewriting the objective and constraints, an equivalent tractable reconstruction is derived. First, we give a distributive algorithm to obtain the Nash equilibrium (NE) of the lower level SUs' noncooperative stochastic game. Thereafter, the stochastic Stackelberg game is discussed under the circumstances that there is no information exchange between PU and SU. Distributed iterative algorithms are designed. Furthermore, a distributive online algorithm is proposed. Finally, simulations are carried out to verify the correctness and demonstrate the effectiveness of proposed algorithms.
\end{abstract}

% \fi
\IEEEpeerreviewmaketitle

\section{Introduction}

Energy harvesting with the capability of scavenging electrical energy from the environment (e.g., solar, ambient radio-frequency (RF) signals) has been a promising maneuver in green communications. Compared to conventional grid energy, the harvested energy has natural green attribute, which makes it extremely suitable to at least partially be the energy source for green networks.
Energy harvesting aided wireless transmission becomes a hot topic in literature. In \cite{IEEETVT15:TianZhang}, we studied the energy harvesting point-to-point communication in cross-layer view. The delay optimal data transmission was investigated when the transmitter has hybrid energy.

Spectrum is another vital resource in wireless communications, besides energy. High spectrum efficiency is a permanent aim in wireless system design. As a mature technique to improve spectrum efficiency, cognitive radio (CR) remains in center area of research. In \cite{IEEETVT14:TianZhang}, we studied the spectrum sharing in OFDM-based cognitive radio networks (CRNs). Analytical as well as iterative hierarchic power allocation algorithms were designed for primary user (PU) and secondary user (SU).

 As a natural idea of combing the virtue of energy harvesting and CR, energy harvesting aided CR has been emerged with the object of lifting the spectrum efficiency with green energy. In \cite{IEEEWCNC16:TianZhang}, we discussed delay optimal data transmission in CR with renewable energy. In \cite{IEEECL16:S. Wu Y. Shin J. Y. Kim and D. I. Kim}, the probability of packet loss was derived by proposing a Markovian battery model for energy harvesting SUs. In \cite{IEEETCOM16:Y. H. Bae and J. W. Baek}, the achievable throughput of energy harvesting SUs in overlay CRNs was analyzed. In \cite{IEEECOMMag15:S. A. R. Zaidi}, the authors proposed a framework to depict the performance of a solar energy harvesting cognitive metro-cellular network.
 In \cite{IEEEICC15:P. He and  L. Zhao}, throughput maximization was studied and optimal algorithm was proposed. In \cite{IEEEJSAC15:Sixing Yin Zhaowei Qu Shufang Li}, the achievable throughput maximization of SU under PU protection was considered, and efficient algorithms were derived. In \cite{IEEEINFOCOM15:S. Gong  L. Duan and P. Wang}, robust power control of energy harvesting SUs to maximize the throughput performance was investigated.
%%%%
%% delay

In this paper, we focus on the cross-layer design of energy harvesting CRNs. The physical layer power allocation is optimized for network layer delay minimization. The main contributions can be concisely stated as three-fold:

 \begin{itemize}
\item A general and practical scenario is studied. The considered aspects include: The intermittence of harvested energy, the randomness of data generation, and the fluctuation of channel state, of each user; The hybrid energy source of PU and the uncertainty of the grid price; The interplay among users and the Signal-Interference-Ratio (SIR) constraint at PU.
\item A stochastic Stackelberg game is formulated in bi-level form. As a basis, Nash equilibrium (NE) of the lower level stochastic sub-game for SUs is computed. Afterwards, the whole stochastic Stackelberg game is investigated under the scenario where no information exchange is possible for PU and SUs.
\item  Based on the theoretical results, efficient distributive off-line and on-line algorithms are designed.
\end{itemize}

\section{System model and Problem formulation}\label{Sec:System model and Problem formulation}
\subsection{System description}

Consider a CR network, where a PU shares spectrum with $N$ SUs. Time is slotted with length $\tau$ each. The PU is denoted as user 0, and SUs are user 1,$\cdots$, user $N$.
Formally, $\mathbb{P}=\{0\}$ denotes the PU set, and $\mathbb{S}=\{1,\cdots,N\}$ is the SU set. Each user is composed of a Tx and a Rx. Each user Tx is equipped with energy harvesting devices, e.g., solar panels. The harvested energy is stored in a battery before usage at each user Tx. In addition, the PU (user 0) can purchase the grid power under an average cost constraint. The grid power with average constraint is applied to alleviate the lability of harvested energy. Then a minimum quality of service (QoS) of the PU can be guaranteed. The grid power price remains static in a slot and fluctuates among slots.
Data are generated from the upper layer of each user and buffered in a first-in-first-out (FIFO) queue. At the beginning of each slot, the user Tx chooses some data from the buffer for delivering to the corresponding Rx. We consider block fading model of the wireless channels, i.e., the channel state remains static during a slot and varies among different slots.
The power gain of the wireless channel between user $i$' Tx and user $j$'s Rx during the $l$-th slot is conveyed as $g_{i,j}[l]$.

Denote the transmitting power of user $i\in \mathbb{P}\cup\mathbb{S}$ at the $l$-th slot as $P_i[l]$, then its error-free instant data rate $R_i[l]$ can be expressed as
\begin{eqnarray}\label{RateinaslotregardingP}
R_i[l]=\log\Big(1+\frac{P_i[l]g_{i,i}[l]}{\sigma^2+\mathcal{I}_i[l]}\Big)
\end{eqnarray}
where $\sigma^2$ is the noise power spectral density at the Rx,
%\begin{eqnarray}
$\mathcal{I}_i[l]=\sum_{j\in \mathbb{P}\cup\mathbb{S},j\ne i} P_j[l]g_{j,i}[l]$
% \end{eqnarray}
is the received interference at user $i$ in the $l$-th slot. We assume unit bandwidth. Generally, the power-rate function can be in other strictly concave and monotonically increasing forms. We utilize
(\ref{RateinaslotregardingP}) as a typical example for better illustration.
Let the stored energy in the battery of user $i$ at the beginning of the $l$-th slot be $E_i[l]$, the harvested energy during the $l$-th slot for user $i$ be $\mathcal{E}_i[l]$. With respect to user $i\in \mathbb{S}$, transmission energy is only allocated from the battery, and then the evolution of the battery energy can be given by
\begin{eqnarray}\label{SUbatteryenergyevolve}
E_i[l+1]=E_i[l]-P_i[l]\tau+\mathcal{E}_i[l]
\end{eqnarray}
with $\mathcal{E}_i[0]$ being the initial battery energy.
Signify the data queue length of user $i$ at the beginning of the $l$-th slot as $Q_i[l]$, the generated data from upper layer during the $l$-th slot for user $i$ as $\mathcal{A}_i[l]$. The data queue length of user $i\in\mathbb{P}\cup\mathbb{S}$ evolves according to
\begin{eqnarray}\label{databufferevolve}
Q_i[l+1]=Q_i[l]-R_i[l]\tau+\mathcal{A}_i[l]
\end{eqnarray}
with $\mathcal{A}_i[0]$ denoting the initial buffer data.

 For conciseness, we define the following notations. $\mathbf{0}$ denotes the vector of \lq\lq all zeros\rq\rq, $\mathbf{E}_i=\big(E_i[1],\cdots,E_i[L]\big)^T$, $\mathbf{R}_i=\big(R_i[1],\cdots,R_i[L]\big)^T$, $\mathbf{Q}_i=\big(Q_i[1],\cdots,Q_i[L]\big)^T$, $\mathbf{P}_i=\big(P_i[1],\cdots,P_i[L]\big)^T$, and $\mathbf{I}_i=\big(\mathcal{I}_i[1],\cdots,\mathcal{I}_i[L]\big)^T$. $T$ is the matrix transpose, $\preceq$ ($\succeq$) means element-wise less (more) than or equal to.

\subsection{Stochastic Stackelberg game formulation}\label{Sec:Stochastic Stackelberg game formulation}

Each user (including the PU and SUs) is rational, selfish, and aims to minimize its own average buffer delay,
 \begin{eqnarray}\label{meandelay}
D_i=\frac{1}{L+1}\sum\limits_{l=1}^{L+1} Q_i[l],
\end{eqnarray}
over the considered $L$ slots.
Meanwhile, to guarantee the PU's transmission, the SIR at the PU should not be less than a constant $\rho > 0$.

To characterize the priority of the PU and competition among users, a Stackelberg game is formulated in bi-level form. The PU is viewed as the leader, and the SUs are as followers.
\subsubsection{Lower level - Noncooperative stochastic game for the SUs}

Define the state of user $i\in\mathbb{S}$ in the $l$-th slot as $$\mathfrak{X}_i[l]=\big(Q_i[l],E_i[l],\{g_{j,i}[l]\}_{j\in\mathbb{P}\cup\mathbb{S}},\mathcal{A}_i[l],\mathcal{E}_i[l]\big)$$ with space $\mathbb{X}_i$. The action of user $i$, $S_i[l]$, is its transmission power, i. e., $S_i[l]=P_i[l]$ with space $\mathcal{S}_i$. Since the SUs should comply with the harvested energy causality and the data causality (i.e., the harvested energy and data can not be utilized or sent before the transmitter receives them, respectively). For state $\mathfrak{X}_i[l]$, the action $S_i[l]=P_i[l]$ should satisfy $0\le P_i[l]\le E_i[l]$ and $0\le R_i[l]\le Q_i[l]$. Let $\mathcal{S}_i(x)$ signify the set of all possible actions of user $i$ when the state is $x\in\mathbb{X}_i$. $\Gamma_{xsy}^i$ means the state transition probability of user $i$. That is to say, if the state of user $i$ is $x\in \mathbb{X}_i$ at a slot and action $s\in\mathcal{S}_i(x)$ is adopted, the next state (state in next slot) of user $i$ is $y\in\mathbb{X}_i$ with probability $\Gamma_{xsy}^i$.
Given the PU's power allocation $\mathbf{P}_0$, the SUs' data transmission can be formulated as a game
$\mathcal{G}=\Big\{\Omega,\mathbb{X},\big\{\mathcal{S}_i\big\}_{i\in\Omega},\big\{\Gamma_{xsy}^i\big\}_{i\in \Omega},\big\{u_i\big\}_{i\in\Omega}\Big\}.$
$\Omega=\mathbb{S}$ is the player set. $\mathbb{X}=\prod\limits_{i\in\Omega}\mathbb{X}_i$ is the state set. The strategy set of user $i$ (SU $i$) $\mathcal{S}_i= \Big\{\mathbf{P}_i: \mathbf{0}\preceq\mathbf{P}_i \preceq \frac{\mathbf{E}_i}{\tau}, \mathbf{R}_i \preceq \frac{\mathbf{Q}_i}{\tau}\Big\}.$ The objective is to minimize the buffer delay, then the utility $u_i=D_i$.

\subsubsection{Upper level - Stochastic optimization of the PU}

The PU has hybrid energy sources, i.e., the harvested energy together with the grid power. Let $\mathbf{V}=\big(V[1],\cdots,V[L]\big)^T$ and $\mathbf{W}=\big(W[1],\cdots,W[L]\big)^T$ with $V[l]$ and $W[l]$ being the allocated grid power and battery power, respectively, in the $l$-th slot. The PU battery energy evolves according to
\begin{eqnarray}\label{PUbatteryenergyevolve}
E_0[l+1]=E_0[l]-W[l]\tau+\mathcal{E}_0[l].
\end{eqnarray}
In addition to the harvested energy causality and buffer data constraints, there are average cost and SIR constraints at the PU. Formally, when the PU can anticipate the SUs¡¯ reactions to its action, the stochastic optimization problem can be given by
\begin{eqnarray}\label{upper level}
\min_{\mathbf{P}_0=\mathbf{V}+\mathbf{W}} u_0=D_0
\end{eqnarray}
\begin{subequations}
\begin{numcases}{\mbox{s.t.}}
\mathbf{0}\preceq\mathbf{W} \preceq \frac{\mathbf{E}_0}{\tau}, \label{HarvestedCausalityCconstraint}\\
\mathbf{R}_0\big(\big\{\mathbf{P}_i^{*}\big\}_{i\in\Omega}\big) \preceq \mathbf{Q}_0,\label{StepRateconstraint}\\
\frac{1}{L} \mathbf{c}^{T}\mathbf{V}\le \mathcal{C},\label{averageCostconstraint}\\
\mathbf{0}\preceq\mathbf{V},\\
\rho \mathbf{I}_0\big(\big\{\mathbf{P}_i^{*}\big\}_{i\in\Omega}\big) \preceq\mathbf{P}_0,\label{ISRConstraint}
\end{numcases}
\end{subequations}
where $\mathbf{c}=(c[1],\cdots,c[L])$ with $c[l]\ge 0$ being the price of the grid power in the $l$-th slot, $\big\{\mathbf{P}_i^{*}\big\}_{i\in\Omega}$ is the NE of the lower stochastic game given $\mathbf{P}_0$.\footnote{ $\big\{\mathbf{P}_i^{*}\big\}_{i\in\Omega}$ is the function of $\mathbf{P}_0$.} $\mathbf{R}_0\big(\big\{\mathbf{P}_i^{*}\big\}_{i\in\Omega}\big)$ and $\mathbf{I}_0\big(\big\{\mathbf{P}_i^{*}\big\}_{i\in\Omega}\big)$ are the rate and received interference for PU, respectively, when PU utilizes $\mathbf{P}_0$ and user (SU) $i$ applies $\mathbf{P}_i^{*}$, $i\in\Omega$. (\ref{HarvestedCausalityCconstraint}) states the causality constraint of the harvested power (energy), (\ref{StepRateconstraint}) is the instant rate constraint due to data causality, (\ref{averageCostconstraint}) means the average cost constraint of the PU, and (\ref{ISRConstraint}) denotes the SIR constraint.

The lower stochastic game together with the upper stochastic optimization constitute the stochastic Stackelberg game. First, the PU chooses a power allocation $\mathbf{P}_0=\mathbf{V}+\mathbf{W}$. Then the SUs run the lower level noncooperative game to attain the NE power allocations given the PU's power accordingly. The PU could anticipate the SUs reactions (i.e., power allocations) to its power allocation. That is to say, the PU knows the SUs' NE power corresponding to each of its transmission power. Then (\ref{upper level}) can be performed at the PU to derive the optimal power $\mathbf{P}_0^{*}$. Once the PU achieves and carries out $\mathbf{P}_0^{*}$, the SUs get the NE power through the lower game given $\mathbf{P}_0^{*}$. The whole play ends thereby.

\emph{Remark: SUs are powered by the rechargeable battery (i.e., the harvested energy) ONLY. In contrast, the PU has hybrid energy supply, i.e., the grid and the rechargeable battery. Corresponding to the stochastic Stackelberg game, the action of an SU is harvested allocation in each slot. And the action of the PU is grid power together with battery allocation.}

% An example of a floating figure using the graphicx package.
% Note that \label must occur AFTER (or within) \caption.
% For figures, \caption should occur after the \includegraphics.
% Note that IEEEtran v1.7 and later has special internal code that
% is designed to preserve the operation of \label within \caption
% even when the captionsoff option is in effect. However, because
% of issues like this, it may be the safest practice to put all your
% \label just after \caption rather than within \caption{}.
%
% Reminder: the "draftcls" or "draftclsnofoot", not "draft", class
% option should be used if it is desired that the figures are to be
% displayed while in draft mode.
%
%\begin{figure}[!t]
%\centering
%\includegraphics[width=2.5in]{myfigure}
% where an .eps filename suffix will be assumed under latex,
% and a .pdf suffix will be assumed for pdflatex; or what has been declared
% via \DeclareGraphicsExtensions.
%\caption{Simulation results for the network.}
%\label{fig_sim}
%\end{figure}

% Note that the IEEE typically puts floats only at the top, even when this
% results in a large percentage of a column being occupied by floats.
%\section{Problem analysis and algorithm design}\label{Sec:Problemanalysis}

\section{Problem reexpression}\label{Sec:Problem reexpression}
For analysis convenience, we reconstruct the objective and constraints.
\subsection{Objective}
According to (\ref{databufferevolve}), $Q_i[l]=Q_i[0]+\sum_{k=1}^{l-1}\big(-R_i[k]\tau+\mathcal{A}_i[k]\big)$. Combining with (\ref{meandelay}), we have
\begin{eqnarray}
D_i&=&Q_i[0]+\frac{1}{L+1}\sum\limits_{l=1}^{L+1}\sum_{k=1}^{l-1}\mathcal{A}_i[k]-\frac{1}{L+1}\sum\limits_{l=1}^{L+1}\sum_{k=1}^{l-1}R_i[k]\tau\nonumber\\
&=&Q_i[0]+\frac{1}{L+1}\sum\limits_{l=1}^{L+1}\sum_{k=1}^{l-1}\mathcal{A}_i[k]-\sum\limits_{l=1}^{L+1} \frac{L+1-l}{L+1}R_i[l]\tau\nonumber\\
\end{eqnarray}
Define
%\begin{eqnarray}\label{discounted average throughput of user i}
$T_i=\sum\limits_{l=1}^{L} \alpha_l R_i[l]\tau$
%\end{eqnarray}
with $\alpha_l=\frac{L+1-l}{L+1}$.
Then $D_i=Q_i[0]+\frac{1}{L}\sum\limits_{l=1}^L\sum_{k=1}^{l-1}\mathcal{A}_i[k]-T_i$. As $Q_i[0]$ and $\mathcal{A}_i[k]$ are physical conditions in the problem,
minimizing $D_i$ is equivalent to maximizing $T_i$. Thereby, the utility can be rewritten as $u_i=T_i$.

\emph{Remark: $T_i$ is the discounted average throughput of user $i$. }

\subsection{Constraints}

For user $i\in \mathbb{S}$, i.e., the SU $i$, $\mathbf{P}_i$ and $\mathbf{R}_i$ are related to $\mathbf{E}_i$ and $\mathbf{Q}_i$ according to (\ref{SUbatteryenergyevolve}) and (\ref{databufferevolve}), respectively. The coupling incurs difficulties in problem analysis.
Denote
 $\bm{\mathcal{E}}_i^a=\big(\mathcal{E}_i^a[1],\cdots,\mathcal{E}_i^a[k],\cdots,\mathcal{E}_i^a[L]\big)^T$ with $\mathcal{E}_i^a[k]=\sum_{l=0}^{k-1}\mathcal{E}_i[l]$,
$\bm{\mathcal{A}}_i^a=\big(\mathcal{A}_i^a[1],\cdots,\mathcal{A}_i^a[k],\cdots,\mathcal{A}_i^a[L]\big)^T$ with $\mathcal{A}_i^a[k]=\sum_{l=0}^{k-1}\mathcal{A}_i[l]$, and
$\mathbf{A}=
\begin{bmatrix}
  1 &  &  &  \\
  1 & 1 &  &  \\
 \vdots &\ddots& \ddots &  \\
  1 & 1 & \cdots & 1 \\
\end{bmatrix}
$.

The strategy set of SU $i$ can be equivalently expressed as
\begin{eqnarray}\label{StrategySet}
\mathcal{S}_i= \Big\{\mathbf{P}_i:\mathbf{A}\mathbf{P}_i\preceq \frac{\bm{\mathcal{E}}_i^a}{\tau}, \mathbf{A}\mathbf{R}_i \preceq \frac{\bm{\mathcal{A}}_i^a}{\tau},\mathbf{0}\preceq\mathbf{P}_i\Big\}.
\end{eqnarray}
 Similarly, for the PU, the constraint can be equivalently rewritten as
\begin{subequations}\label{PUconstraints}
\begin{numcases}{}
\mathbf{A}\mathbf{W} \preceq \frac{\bm{\mathcal{E}}_0^a}{\tau}, \label{HarvestedCausalityCconstraintqui}\\
\mathbf{0}\preceq\mathbf{W},\\
\mathbf{A}\mathbf{R}_0\big(\big\{\mathbf{P}_i^{*}\big\}_{i\in\Omega}\big) \preceq \frac{\bm{\mathcal{A}}_0^a}{\tau},\label{StepRateconstraintqui}\\
\frac{1}{L} \mathbf{c}^{T}\mathbf{V}\le \mathcal{C},\label{averageCostconstrainteqi}\\
\mathbf{0}\preceq\mathbf{V},\\
\rho \mathbf{I}_0\big(\big\{\mathbf{P}_i^{*}\big\}_{i\in\Omega}\big) \preceq\mathbf{P}_0,\label{ISRConstrainteqi}
\end{numcases}
\end{subequations}
%where $\mathbf{W}^a=(W^a[1],\cdots,W^a[k],\cdots,W^a[L])$ with $W^a[k]=\sum_{l=1}^k W[l]$ , $\mathbf{R}_0^a\big(\big\{\mathbf{P}_i^{*}\big\}_{i\in\Omega}\big)$ is the $\mathbf{R}_0^a$ given $\big\{\mathbf{P}_i^{*}\big\}_{i\in\Omega}$ and $\mathbf{P}_0$.

\emph{Remark: In the equivalently reconstructed constraints (\ref{StrategySet}) and (\ref{PUconstraints}), $\bm{\mathcal{E}}_i^a$ and $\bm{\mathcal{A}}_i^a$ are accumulated harvested energy and data arrivals. Then there are only the physical conditions. Consequently, the decoupling completes.}

\section{Game analysis and algorithm proposal}\label{Sec:Problemanalysis}
%%%%%%%
\theoremstyle{definition} \newtheorem{definition}{Definition}
\theoremstyle{lemma} \newtheorem{lemma}{Lemma}

%%%%%%%

The stochastic Stackelberg game is formulated in bilevel manner. Accordingly, the algorithm to get the NE of the lower level problem is proposed first, and then distributed iterative algorithms are proposed for the upper level (whole game), Specially, an online scheme is given.

\subsection{On lower level SUs' stochastic game}

%Concerning the existence of the NE for the lower level game, we have the following lemma.
%\begin{lemma}\label{Lemma:NE existence}
%The NE of the lower level SUs' game, $\mathcal{G}$, exists.
%\end{lemma}
%\begin{IEEEproof}
%See Appendix \ref{Proof:NE existence}.
%\end{IEEEproof}
%Define $\mathbf{P}=\big\{\mathbf{P}_i\}_{i\in\mathbb{S}}$, $\mathbf{T}=\big(T_1,\cdots,T_N\big)^T$ with $T_i$ being expressed by (\ref{discounted average throughput of user i}). $\mathbf{T}$ is function of $\mathbf{P}$, and is expressed as $\mathbf{T}(\mathbf{P})$ thereby. The uniqueness of $\mathcal{G}$ is stated as
%\begin{lemma}\label{Lemma:NE uniqueness}
%Signify $g(\mathbf{P},\mathbf{r})$ as the pseudogradient of $\psi(\mathbf{P},\mathbf{r}):=\mathbf{r}^T\mathbf{T}(\mathbf{P})$ with $\mathbf{r}=(r_1,\cdots,r_N)^T\succeq \mathbf{0}$. If $G(\mathbf{P},\mathbf{r})+G(\mathbf{P},\mathbf{r})^T$ is negative definite with respect to $\mathbf{P}$, in which $G(\mathbf{P},\mathbf{r})$ is the the Jacobian with respect to $\mathbf{P}$, then $\mathcal{G}$  has a unique NE.
%\end{lemma}
%\begin{IEEEproof}
%See Appendix \ref{Proof:NE uniqueness}.
%\end{IEEEproof}
%
%\emph{Remark: Lemma \ref{Lemma:NE uniqueness} gives a sufficient condition for the NE uniqueness. Other sufficient conditions can be referred to previous related works (e. g., \cite{IEEETVT14:TianZhang}\cite{IEEESP08:G. Scutari D. P. Palomar and S. Barbarossa} and references therein). In the paper, we investigate the unique NE situation ONLY.}
%
We focus on \emph{\lq\lq How to compute the NE of lower game?\rq\rq}
The optimization problem of user $i$ (given PU and other SUs' transmissions) can be expressed as
\begin{eqnarray}\label{OptProblemofuseriregardingP}
\max_{\mathbf{P}_i} u_i=T_i
\end{eqnarray}
\begin{subequations}
\begin{numcases}{\mbox{s.t.}}
\mathbf{A}\mathbf{P}_i\preceq \frac{\bm{\mathcal{E}}_i^a}{\tau},\\
\mathbf{0}\preceq \mathbf{P}_i,\\
\mathbf{A}\mathbf{R}_i \preceq \frac{\bm{\mathcal{A}}_i^a}{\tau}
\end{numcases}
\end{subequations}

Through solving (\ref{OptProblemofuseriregardingP}), we derive the best response of SU $i$ given the PU and other SUs' actions. Thereafter, we have the following off-line distributive iterated algorithm (Algorithm 1) for obtaining the NE of the lower game.

\emph{Remark: In Algorithm 1, user $i$ requires its own channel state sequence $\{g_{ii}[l]\}_{l=1,\cdots,L}$, data arrival sequence $\{\mathcal{A}_i[l]\}_{l=1,\cdots,L}$, harvested energy sequence $\{\mathcal{E}_i[l]\}_{l=1,\cdots,L}$, and measures the sequence of aggregated interference from other users $\mathbf{I}_i$. Hence, Algorithm 1 can be distributively applied.
In addition, Algorithm 1 requires all-slot (past, current and future) system
information (e.g., $\{g_{ii}[l]\}_{l=1,\cdots,L}$, $\{\mathcal{A}_i[l]\}_{l=1,\cdots,L}$,
$\{\mathcal{E}_i[l]\}_{l=1,\cdots,L}$, and $\mathbf{I}_i$) as a priori, it is an off-line algorithm}
 \begin{table}[]
 %\caption{\label{table 1} Algorithm}
 \centering
 \begin{tabular}{lcl}
  \toprule
  \textbf{Algorithm 1: Iterated distributive off-line Algorithm }\\
  ~~~~~~~~~~~~\textbf{for deriving NE solution of lower-level game}\\
  \midrule
 Step 1:  $k=0$, initialize feasible policy for $N$ SUs, $\mathbf{P}_i^0$.\\
 Step 2: For every $i \in \Omega$, Update $\mathbf{P}_i^{k+1}$ as the optimal policy (best \\
 ~~~~response) of user $i$ by solving (\ref{OptProblemofuseriregardingP}), given $\mathbf{P}_{-i}^k:=\big\{\mathbf{P}_{j}^k\big\}_{j\in\Omega,j\ne i}$.  \\
 Step 3: $k=k+1$, go to Step 2 until convergence. \\
  \bottomrule
 \end{tabular}
\end{table}

\subsection{On upper level \& the whole stochastic Stackelberg game}

When information exchange is possible and the PU is assumed to have accessibility of the private information of SUs (each SU's system state, e.g., data arrival, energy arrival, etc.), it could anticipate the reactions of the SUs (followers). (\ref{upper level}) can be mathematically carried out at the PU.
The stochastic Stackelberg game is generally a bilevel programming problem\cite{AOR07:B. Colson P. Marcotte G. Savard}, and specifically belongs to mathematical programs with equilibrium constraints (MPEC)\cite{Book96:Z.-Q. Luo J.-S. Pang and D.Ralph}. However, in the scenario that no information exchange exists among PU and SUs, the PU could not obtain the SUs' information, and it can not anticipate SUs' precise reactions to its action thereby. Accordingly, the PU's upper level problem (\ref{upper level}) can not exactly execute. Stackelberg equilibrium (SE) is unavailable. \emph{How to investigate the problem in this scenario} becomes a natural topic.
 We propose an iterative algorithm referred to as the NE based re-Optimization (NEO) algorithm (in Table \ref{Algorithm:NEO}) to give a solution under this situation. The NEO algorithm is suboptimal to SE.

In Step 1, an initial PU's power allocation is set. Next, in Step 2, given the PU's power, SUs run the lower level game to arrive at the NE state. In Step 3, for fixed SUs' power allocations, the PU's optimization problem (\ref{upper level}) becomes
\begin{eqnarray}\label{PU's problem given SUs' power}
\max_{\mathbf{P}_0=\mathbf{V}+\mathbf{W}} T_0
\end{eqnarray}
\begin{subequations}\label{constraints of PU's problem given SUs' power}
\begin{numcases}{\mbox{s.t.}}
\mathbf{A}\mathbf{W} \preceq \frac{\bm{\mathcal{E}}_0^a}{\tau}, \label{}\\
\mathbf{0}\preceq\mathbf{W},\\
\frac{1}{L} \mathbf{c}^{T}\mathbf{V}\le \mathcal{C},\label{}\\
\mathbf{0}\preceq\mathbf{V},\\
\mathbf{A}\mathbf{R}_0 \preceq \frac{\bm{\mathcal{A}}_0^a}{\tau},\label{RateConsInPrivateUnavailable}\\
\rho \mathbf{I}_0 \preceq\mathbf{P}_0,\label{SIRConsInPrivateUnavailable}
\end{numcases}
\end{subequations}

%\emph{Remark: (\ref{SIRConsInPrivateUnavailable}) can be equivalently stated as $R_0[l]\ge \log\big(1+\frac{\rho \mathcal{I}_0[l]g_{0,0}[l]}{\sigma^2+\mathcal{I}_0[l]}\big)$. Then $\sum_{l=1}^kR_0[l] (i.e., R_0^ a[k])\ge \sum_{l=1}^k\log\big(1+\frac{\rho \mathcal{I}_0[l]g_{0,0}[l]}{\sigma^2+\mathcal{I}_0[l]}\big)$ should hold. On the other hand, (\ref{RateConsInPrivateUnavailable}) means $R_0^ a[k]\le \mathcal{A}_0^a[l]$.
%Once
%\begin{eqnarray}\label{SufficientCondiForNonsolution}
%\mathcal{A}_0^a[l] <\log\big(1+\frac{\rho \mathcal{I}_0[l]g_{0,0}[l]}{\sigma^2+\mathcal{I}_0[l]}\big),
%\end{eqnarray}
%(\ref{RateConsInPrivateUnavailable}) and (\ref{SIRConsInPrivateUnavailable}) can not satisfied simultaneously. That is to say, there are no solutions for (\ref{PU's problem given SUs' power}).
%(\ref{SufficientCondiForNonsolution}) demonstrates a sufficient condition of the no solutions. Since the data generation of the PU is random, the probability of the case that few data available in the buffer is not zero. In this sense, SIR in average is more practical.}

By deriving the optimal solution of (\ref{PU's problem given SUs' power}), we get a renewed PU power. Given the updated PU power, we get a renewed SUs' NE... By repeating the process, a steady state, which is the algorithm output, can be arrived in the end.

  \begin{table}[!t]
 \caption{\label{Algorithm:NEO}}
 \centering
 \begin{tabular}{lcl}
  \toprule
  \textbf{NEO Algorithm}\\
  \midrule
 Step 1: $k=0$, initialize a power allocation for the PU, $\mathbf{P}_0^0$.\\
 Step 2: Given $\mathbf{P}_0^k$, deriving the NE of the SUs' game as $\big\{\mathbf{P}_i^{k}\big\}_{i\in\Omega}$\\
 ~~~~~~~~~through Algorithm 1.\\
 Step 3: Given $\big\{\mathbf{P}_i^{k}\big\}_{i\in\Omega}$, obtain $\mathbf{P}_0^*$ as the optimal solution of (\ref{PU's problem given SUs' power}).  \\
  ~~~~~~~~~Update
 $\mathbf{P}_0^{k+1}=\eta\mathbf{P}_0^*+(1-\eta)\mathbf{P}_0^{k}$, $0<\eta\le 1$ is a step size.\\
 Step 4: $k=k+1$, go to Step 2 until convergence. \\
  \bottomrule
 \end{tabular}
\end{table}

\emph{Remark: On one hand, the PU needs its own information (e.g., $\bm{\mathcal{E}}_0$, $\bm{\mathcal{A}}_0$, etc.) and measures the received aggregated interference $\mathbf{I}_0$ in solving (\ref{PU's problem given SUs' power}). On the other hand, Algorithm 1 is a distributive scheme. Hence NEO algorithm is a distributive approach.}

% Note that the IEEE does not put floats in the very first column
% - or typically anywhere on the first page for that matter. Also,
% in-text middle ("here") positioning is typically not used, but it
% is allowed and encouraged for Computer Society conferences (but
% not Computer Society journals). Most IEEE journals/conferences use
% top floats exclusively.
% Note that, LaTeX2e, unlike IEEE journals/conferences, places
% footnotes above bottom floats. This can be corrected via the
% \fnbelowfloat command of the stfloats package.
\subsection{An on-line algorithm}

In above analysis and algorithm design, we assume the information over all considered $L$ slots as a prior (i.e., off-line case). In practical applications, off-line condition is hard to satisfy if not impossible (or the accomplishment cost is too high). Then, online scheme, which requires only current slot information by contrast, is important and necessary.

In each slot, the PU and SUs play a Stackelberg game, where the strategy is power allocation in current slot and the utility is the instant rate with a coefficient.

The SUs form a noncooperative game in each slot. Formally, in the $l$-th slot, for user $i\in\mathbb{S}$

\begin{eqnarray}\label{OptProbForSUiinOneSlotGame}
\max_{P_i[l]} \alpha_l R_i[l]\tau
\end{eqnarray}
\begin{subequations}
\begin{numcases}{\mbox{s.t.}}
0\le P_i[l]\le \frac{E_i[l]}{\tau},\\
0\le R_i[l]\le \frac{Q_i[l]}{\tau}.
\end{numcases}
\end{subequations}
The solution (i.e., best response of SU $i$) is
\begin{eqnarray}\label{bestResponsofSUiInOneslotGame}
P_i^*[l]=\min\bigg\{\frac{E_i[l]}{\tau},\big(e^{Q_i[l]/\tau}-1\big)\frac{\sigma^2+\mathcal{I}_i[l]}{g_{i,i}[l]}\bigg\}.
\end{eqnarray}

\emph{Remark: (\ref{bestResponsofSUiInOneslotGame}) means transmitting as many data as possible in each slot, i.e., greedy policy.}

Denote the available grid power budget at the beginning of the $l$-th slot as $\mathcal{B}[l]$ with $\mathcal{B}[1]=L\mathcal{C}$. The budget evolution is
\begin{eqnarray}
\mathcal{B}[l+1]=\mathcal{B}[l]-c[l] V[l].
\end{eqnarray}
For the PU, as the grid power cost constraint is in average sense. In the online design, we uniformly allocate the budget constraint (formally, (\ref{uniform budget constraint})).
In the $l$-th slot, the PU's problem becomes
\begin{eqnarray}\label{OptProbForPUinOneSlotGame}
\max_{P_0[l]=V[l]+W[l]} \alpha_lR_0[l]\tau
\end{eqnarray}
\begin{subequations}
\begin{numcases}{\mbox{s.t.}}
0\le W[l]\le \frac{E_0[l]}{\tau},\\
0\le c[l]V[l]\le \frac{\mathcal{B}[l]}{L-l+1},\label{uniform budget constraint}\\
0\le R_0[l]\le \frac{ Q_0[l]}{\tau},\\
\rho\mathcal{I}_0[l]\le P_0[l].
\end{numcases}
\end{subequations}

The solution is
\begin{eqnarray}\label{bestResponsofPUInOneslotGame}
%\lefteqn{
P_0^*[l]&=&\min\bigg\{\frac{E_0[l]}{\tau}+\frac{\mathcal{B}[l]}{(L-l+1)c[l]}, \nonumber\\
%}\nonumber\\
&&\big(e^{Q_0[l]/\tau}-1\big)\frac{\sigma^2+\mathcal{I}_0[l]}{g_{0,0}[l]}\bigg\}.
\end{eqnarray}

\emph{Remark: (\ref{bestResponsofPUInOneslotGame}) demonstrates that the best response of the PU is utilizing the greedy policy in each slot.
%It is assumed that $\min\bigg\{\frac{E_0[l]}{\tau}+\frac{\mathcal{B}[l]}{(L-l+1)c[l]},\big(e^{Q_0[l]}-1\big)\frac{\sigma^2+\mathcal{I}_0[l]}{g_{0,0}[l]}\bigg\}\ge \rho \mathcal{I}_0[l]$. Otherwise, there is no solution of (\ref{OptProbForPUinOneSlotGame}).
}

When obtaining $P_0^*[l]$, the grid power and harvested energy
allocations are given as follows: If $P_0^*[l]\le \frac{\mathcal{B}[l]}{(L-l+1)c[l]}$, $V[l]=P_0^*[l]$ and $W[l]=0$;
Otherwise, $V[l]=\frac{\mathcal{B}[l]}{(L-l+1)c[l]}$ and $W[l]=P_0^*[l]-\frac{\mathcal{B}[l]}{(L-l+1)c[l]}$.

%\emph{Remark: In GoG, the remaining grid power budget is averaged over rest of slots as constraint. In contrast, the remaining renewable has no such inefficiency. Thus, the grid power budget is utilized in priority.}

The iterative online algorithm, Greedy one-slot Game (GoG) algorithm, is outlined in Table \ref{Algorithm:GoG}.
  \begin{table}[!t]
 \caption{\label{Algorithm:GoG}}
 \centering
 \begin{tabular}{lcl}
  \toprule
  \textbf{GoG Algorithm}\\
  \midrule
 Step 1: $k=0$, initialize a power allocation for the PU, $P_0^0$.\\
 Step 2: (Obtaining the NE of SUs given $P_0^k$)\\
 ~~~Step 2-1: $m=0$, set initial power allocations of $N$ SUs $\{P_i^0\}_{i\in\mathbb{S}}$.\\
 ~~~Step 2-2: Renew $P_i^{m+1}$ utilizing (\ref{bestResponsofSUiInOneslotGame}) given $P_{-i}^m:=\{P_{j}^m\}_{j\in\mathbb{S},j\ne i}$ \\
 ~~~~~~~~~~~~~~and $P_0^k$ for $i\in\mathbb{S}$\\
 ~~~Step 2-3: $m=m+1$, go to Step 2-2 until convergence (or stopping \\
~~~~~~~~~~~~~~condition holds). The final output is $\big\{P_i^{k}\big\}_{i\in\mathbb{S}}$.\\
 Step 3: Given $\big\{P_i^{k}\big\}_{i\in\mathbb{S}}$, update $P_0^{k+1}$ applying (\ref{bestResponsofPUInOneslotGame}). \\
 Step 4: $k=k+1$, go to Step 2 until convergence. \\
  \bottomrule
 \end{tabular}
\end{table}
In a slot, we apply GoG algorithm to get the SUs' power and PU's power (including the grid and harvested). Update related system information (e.g., the system state, the grid power budget), and move to next slot until completing all slots.

\emph{Remark: In GoG algorithm, each user requires its own information and gauges the received aggregated inference. Then GoG can be utilized distributively.}

\section{Numerical results}\label{Sec:Numerical results}
In the section, simulations are performed to demonstrate the correctness and effectiveness of proposed algorithms. The noise power spectral density $\sigma^2=0.1$. The time slot
length $\tau=1$.
\begin{figure}[]
\centering
\subfigure[Power allocation of SU 1]{\includegraphics[width=3.3in]{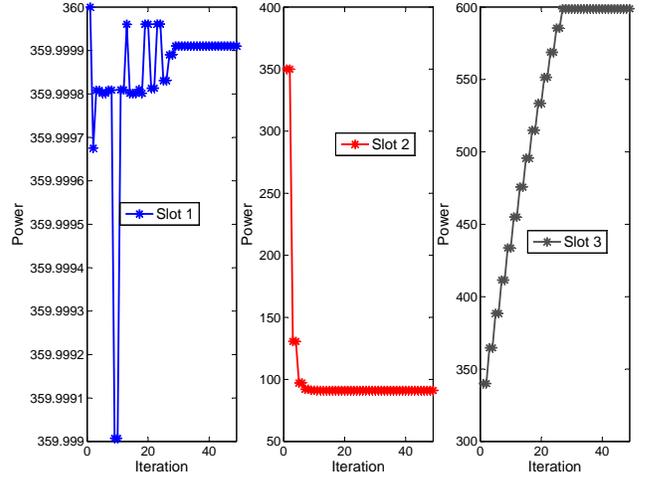}}
\subfigure[Power allocation of SU 2]{\includegraphics[width=3.2in]{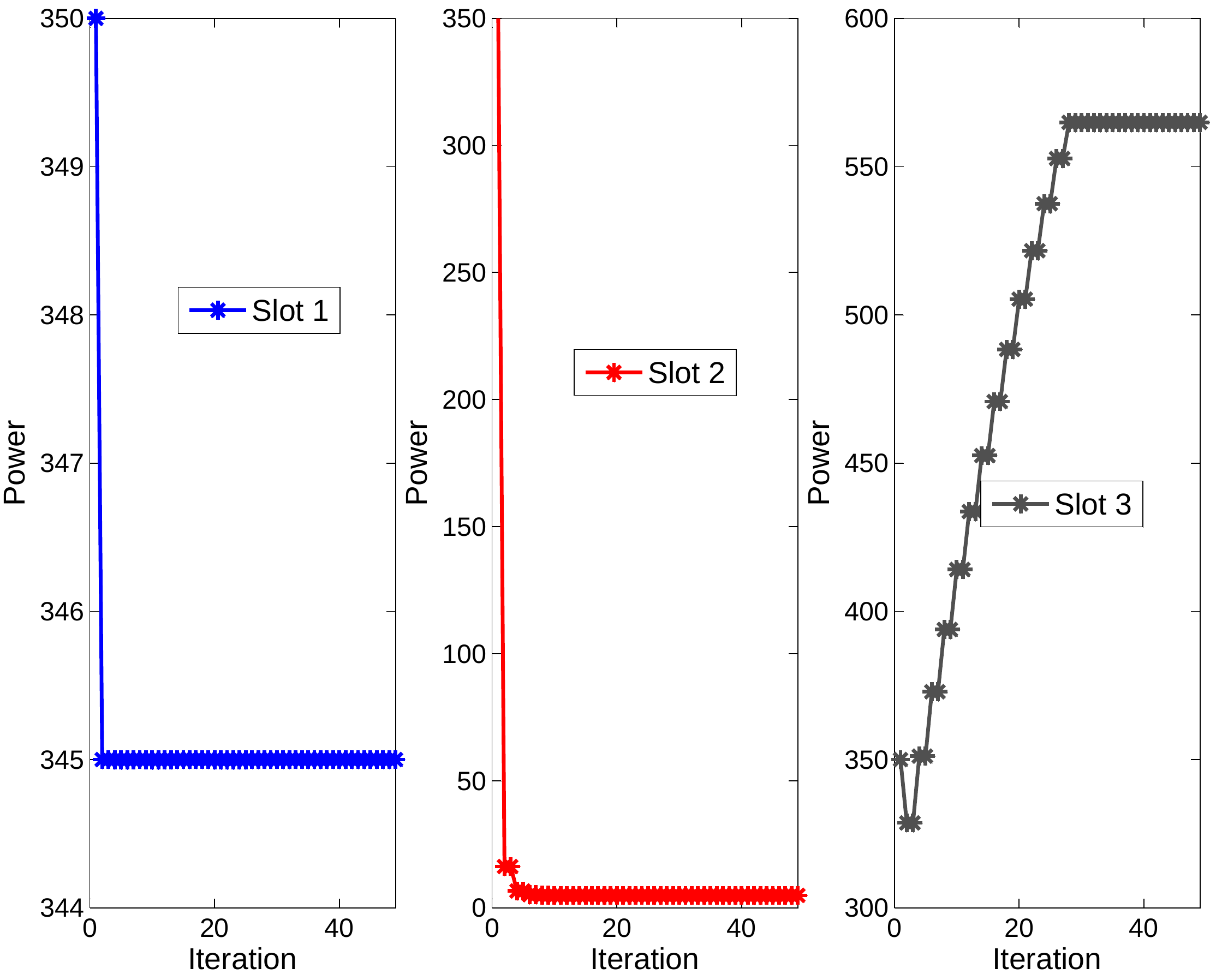}}
\caption{NE power of SUs given $\mathbf{P}_0=(20,20,20)^T$ when Algorithm 1 is applied}
\label{Sim:1}
\end{figure}
%\subsection{1-2-3 setting}
In the settings, 1 PU, 2 SUs and 3 slots are considered, i.e., $N=2$ and $L=3$.

%

%\begin{figure}[t]
%\centering
%\subfigure[Power allocation of SU 1]{\includegraphics[width=3.1in]{Sim2a.eps}}
%\subfigure[Power allocation of SU 2]{\includegraphics[width=3.0in]{Sim2b.eps}}
%\caption{NE power of SUs given $\mathbf{P}_0=(100,100,100)^T$ with Algorithm 1}
%\label{Sim:2}
%\end{figure}

Fig.\ref{Sim:1} draws the SUs' NE power given the PU's power $(100,100,100)^T$. In the simulations, Algorithm 1 is applied. The initial power are $\mathbf{P}_1^0=(360,350,340)^T$ and
$\mathbf{P}_2^0=(350,350,350)^T$. The channel states are set as
$g_{0,1}=(0.09,0.07,0.06)^T$, $g_{0,2}=(0.05,0.1,0.08)^T$, $g_{1,1}=(0.1,0.15,0.13)^T$,
$g_{1,2}=(0.07,0.11,0.085)^T$, $g_{2,2}=(0.12,0.14,0.16)^T$, and $g_{2,1}=(0.07,0.07,0.08)^T$. The data arrival $\mathcal{A}_1=(1,2,1)$ and
$\mathcal{A}_2=(1,0,1)$. The renewable energy arrival $\bm{\mathcal{E}}_1=(360,350,340)$ and $\bm{\mathcal{E}}_2=(345, 380,370)$. $\mathcal{A}_i:=(\mathcal{A}_i[0],\mathcal{A}_i[1],\mathcal{A}_i[2])$ ($i=1,2,3$). $\bm{\mathcal{E}}_i:=(\mathcal{E}_i[0],\mathcal{E}_i[1],\mathcal{E}_i[2])$ ($i=1,2,3$).
From the figures, we can see that Algorithm 1 reaches the convergent points (i.e., NE) after not many iterations.

\begin{figure}[]
\centering
\subfigure[ $\eta=0.9$]{\includegraphics[height=2.5in,width=3.2in]{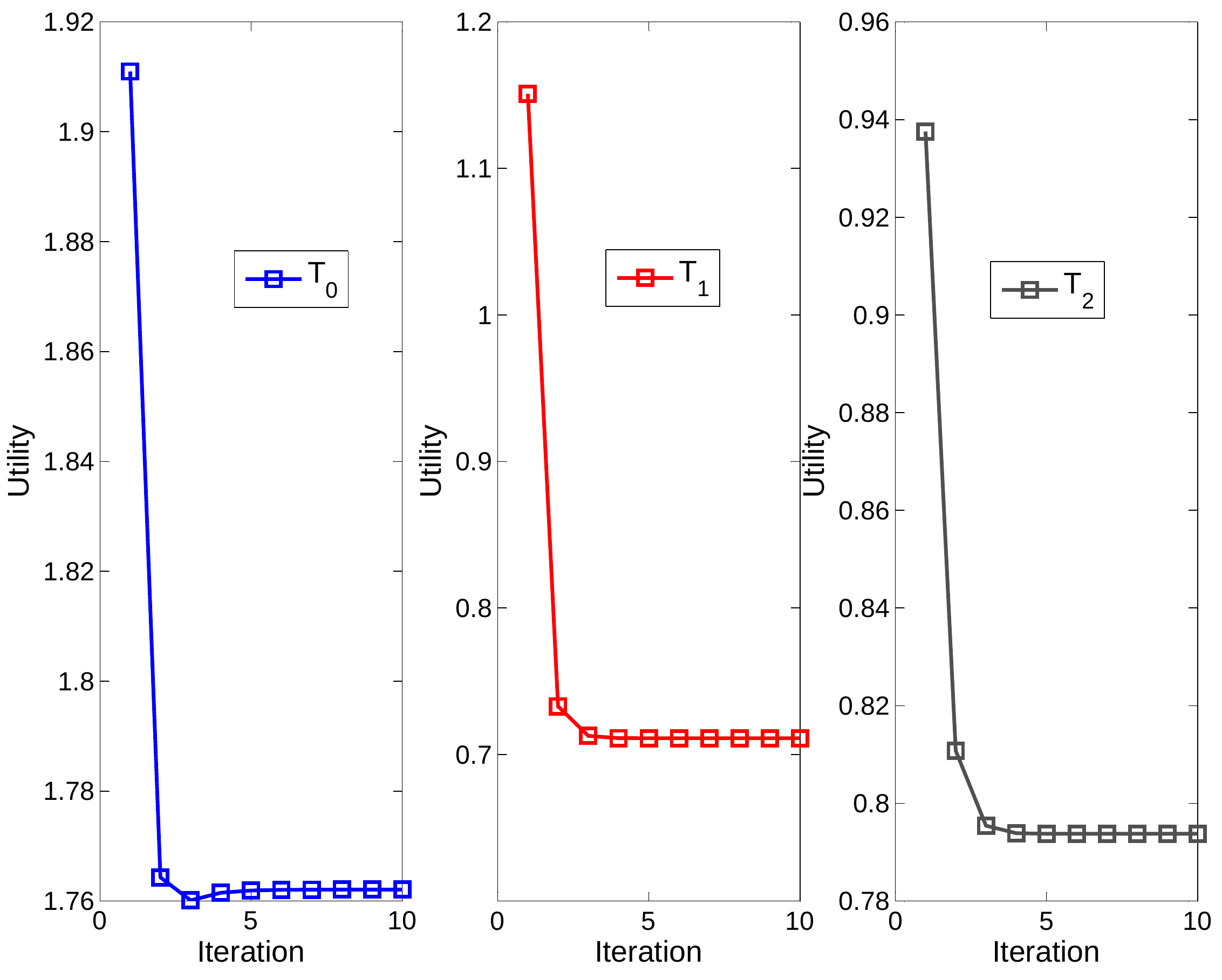}}
%\subfigure[$\eta=0.5$]{\includegraphics[height=2.1in,width=3.0in]{Sim3b.eps}}
\subfigure[ $\eta=0.3$]{\includegraphics[height=2.5in,width=3.2in]{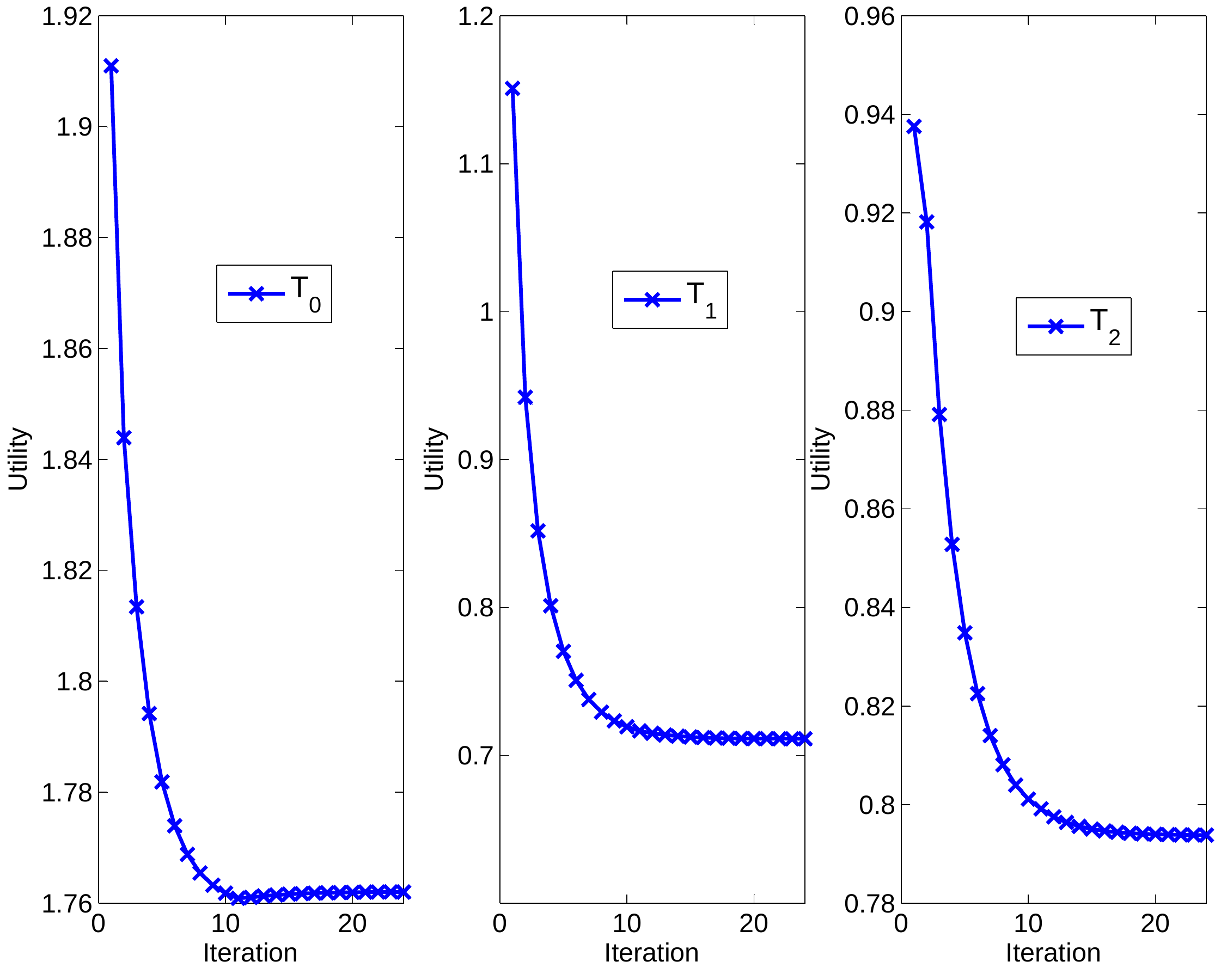}}
%\subfigure[$\eta=0.1$]{\includegraphics[height=2.1in,width=3.0in]{Sim3d.eps}}
\caption{NEO algorithm with different stepsize $\eta$}
\label{Sim:3}
\end{figure}

Fig. \ref{Sim:3} illustrates the utility performance of NEO algorithm. Different values of stepsize, $\eta=0.9$ and $\eta=0.3$, are set for comparisons. The initial PU power is $(300, 200, 200)^T$.
The channel states are set as $g_{0,0}=(0.2,0.3,0.12)^T$
$g_{0,1}=(0.09,0.07,0.06)^T$, $g_{0,2}=(0.05,0.1,0.08)^T$, $g_{1,1}=(0.1,0.15,0.13)^T$, $g_{1,0}=(0.1,0.06,0.06)^T$,
$g_{1,2}=(0.07,0.11,0.085)^T$, $g_{2,2}=(0.12,0.14,0.16)^T$, $g_{2,1}=(0.07,0.07,0.08)^T$ and $g_{2,0}=(0.09,0.065,0.08)^T$. $\rho=0.01$. The data arrival $\mathcal{A}_0=(3,5,8)$, $\mathcal{A}_1=(1,2,1)$ and
$\mathcal{A}_2=(1,0,1)$. The harvested energy arrival $\bm{\mathcal{E}}_0=(600, 500,450)$, $\bm{\mathcal{E}}_1=(360,350,340)$ and $\bm{\mathcal{E}}_2=(345, 380,370)$. $\mathbf{c}=(1, 1.2, 0.9)^T$ and $\mathcal{C}=100$. We can observe that when $\eta$ decreases, the convergence speed becomes slow. This can be explained as follows: In the update of PU power (Step 3),the larger $\eta$ is , the nearer the renewed power approaches the best response. Generally, best response is the shortest path to the final points. Hence larger $\eta$ leads to higher speed.
The NEO algorithm finally converges to $T_0=1.7620$, $T_1= 0.7111$ and $T_2= 0.7938$.

\begin{figure}[]
\centering
\includegraphics[width=3.3in]{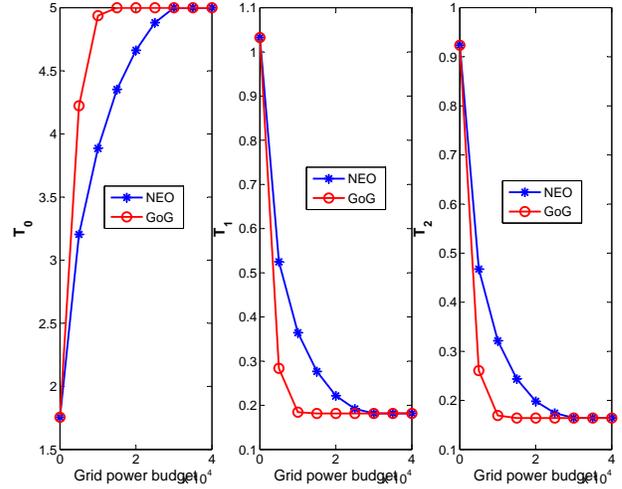}
%\subfigure[Regarding PU data arrival scale, $f$ ]{\includegraphics[width=3.0in]{Sim5.eps}\label{Sim:5}}
% via \DeclareGraphicsExtensions.
\caption{Utility performance of NEO and GoG regarding $\mathcal{C}$}
\label{Sim:45}
\end{figure}

Fig. \ref{Sim:45} plots the utility performance of NEO and GoG with respect to the grid power budget $\mathcal{C}$. The channel gains are
$g_{0,0}=(0.2,0.3,0.12)^T$
$g_{0,1}=(0.09,0.07,0.06)^T$, $g_{0,2}=(0.05,0.1,0.08)^T$, $g_{1,1}=(0.2,0.15,0.23)^T$, $g_{1,0}=(0.1,0.06,0.04)^T$,
$g_{1,2}=(0.06,0.10,0.08)^T$, $g_{2,2}=(0.12,0.2,0.18)^T$, $g_{2,1}=(0.07,0.09,0.06)^T$ and $g_{2,0}=(0.07,0.06,0.08)^T$. $\rho=0.01$. The step size of NEO is $\eta=0.8$. The energy arrival $\bm{\mathcal{E}}_0=(600, 500,450)$, $\bm{\mathcal{E}}_1=(350,400,340)$ and $\bm{\mathcal{E}}_2=(345, 380,350)$
The data arrival $\mathcal{A}_1=(2,2,1)$ and
$\mathcal{A}_2=(1,3,2)$. The grid power price $\mathbf{c}=(1.5, 2.0, 0.9)^T$. $\mathcal{A}_0=(3,4,3)$.
 By comparing the NEO and GoG, we can see that the NEO has better SU performance and the GoG has better PU performance. From the figure, it can be observed that the PU's utility increases and the SUs' utilities decrease at first, and all remain constant then when we increase the grid power budget $C$. The explanations are as follows: When $\mathcal{C}$ is mild, the increase will give more power for PU transmission, then the utility performance of PU improves. More PU transmission produces more interference to SUs, and the SUs' utility performance degrades accordingly. Once $\mathcal{C}$ is larger than some value (e.g., $30000$ in the figure), the PU utility reaches as the upper bound $3*\frac{3}{4}+4*\frac{2}{4}+3*\frac{1}{4}=5$. After that, the instant rate constraint becomes active and the utility remains static. Meanwhile, the interference from PU becomes static at SUs and utility becomes some constant.

\section{Conclusion}\label{Sec:conclusion}

Energy harvesting aided CRN is investigated in cross-layer view. We study the physical layer power control for the sake of minimizing the network delay. A stochastic Stackelberg game is constructed in bilevel format. Distributive off-line algorithms, NEO algorithms, and distributive on-line algorithm, GoG algorithm, are proposed based on theoretical analyses. Numerical results demonstrate the effectiveness of proposed algorithms.


%% use section* for acknowledgment
%\section*{Acknowledgment}



%\bibliographystyle{IEEEtran}

\end{document}